\documentclass[aps, preprint, showpacs]{revtex4}
\begin{document}
\title{Nature of the Angular Momentum of Light: Rotational Energy and Geometric Phase}
\author { S. C. Tiwari\\
Institute of Natural Philosophy\\
c/o 1 Kusum Kutir Mahamanapuri,Varanasi 221005, India}
\begin{abstract}
 The nature of intrinsic/extrinsic character
of angular momentum is defined in terms of the kind of the associated 
rotational energy of the light. The salient features of the spin energy of light and
photon are highlighted. Spin angular momentum is intrinsic while orbital angular
momentum possesses quasi-intrinsicness only if the vortex-like singularities are present.
The claimed spin to orbital angular momentum conversion is interpreted in terms of spin redirection geometric phase. It is pointed out that this interpretation validates our angular momentum holonomy conjecture.
\end{abstract}
\maketitle

1. Rotation seems to have an absoluteness; in the case of optics the Sagnac effect
discovered in 1913 \cite{1} indicates this absoluteness; for review on recent controversies and numerous explanations see \cite{2}. In another context spin offers a profound example 
for the pure rotation. Poynting in 1909 associated angular momentum (AM) to radiation, 
and Beth in 1935 reported first direct
detection of spin angular momentum (SAM) carried by circularly polarized light using
doubly refracting waveplates \cite{3}. At that time it had become known in quantum theory
that photon possessed SAM of ${\pm\hbar}$. AM calculated using the Poynting
vector for radiation has another component, namely the orbital angular momentum (OAM)
besides SAM. Is OAM intrinsic just like SAM? This question has been discussed recently in this journal, and has acquired renewed significance in the light of recent advances
in lasers, quantum optics, and quantum information science. Spin for single photon has fundamental significance more so when the word 'photon' is often a 
shorthand notation for energy $h\nu$, and there exist unresolved conceptual problems in ascribing ontological reality to it \cite{4}.

A typical standard example of OAM is that of multipole radiation field 
such that $m\hbar$ units of z-component of OAM per photon of energy $\hbar\omega$ could 
be associated in a quantum interpretation \cite{5}. Relatively recently Allen et al
\cite{6} recognised that Laguerre-Gaussian (LG) laser modes possessed well defined
OAM of $l\hbar$ per photon. The azimuthal index $l$ determines angular phase dependence 
of the form $exp(il\phi)$. Heuristically analogy with the dislocations 
and disclinations in crystals could be made to describe the singularities in
a general monochromatic electromagnetic wavefield \cite{7}. In this approach LG
beam has a phase dislocation on the beam axis and the index $l$ acquires a topological
significance-apparently OAM would be intrinsic. Experiments \cite{8,9} confirm that both
SAM and OAM from light beams are transferred to small particles as expected from theoretical analysis. It is suggested in \cite{9} that OAM could be either intrinsic
or extrinsic. A recent paper \cite{10} explores the quasi-intrinsic features of OAM. In
an important experiment \cite{11} conversion of SAM to OAM has been claimed using an
inhomogeneous anisotropic optical element called q-plate. Obviously physical mechanism
to understand this counter-intuitive process and the intrinsic or extrinsic character
of the AM of light would give new insights into the nature of light and
physics of crystal optics. In the present Letter we address these issues.

Perusal of the published literature shows that the energy associated with pure rotation
or spin for light and photon remains obscure. First we explain the meaning of this
statement and attempt to bring out its significance as regards to the debate on the
intrinsic or extrinsic nature of SAM and OAM for light.

An important subject in modern optics is that of geometric phases.
It was conjectured in \cite{12} that AM exchange was the physical
mechanism responsible for the geometric phases; see recent elucidation in \cite{13}. 
Reported observations in
\cite{11} are analyzed in the perspective of geometric phase in the present paper. We propose that the induced geometric circular birefringence \cite{14} in addition to normal birefringence determines the optical properties of the q-plates. If this interpretation turns out to be valid
Marrucci et al experiment would be the first direct proof of our conjecture made in
\cite{12,13}; variants on this experiment are also suggested.

2. Mechanical equivalence of electromagnetic fields is embodied in the energy, momentum,
and angular momentum of the fields and their conservation laws, and constitute the 
standard textbook matter, e.g. \cite{5}. In an illuminating monograph \cite{15} formal
technical details and insightful remarks can be found; we omit formal details. Let us 
summarize the salient features of the conceptual problems related with the AM. In the relativistic covariant formulation the action integral is constructed
from the Lagrangian density
\begin{equation}
L= -(1/16\pi) F^{\mu\nu} F_{\mu\nu}
\end{equation}
where the electromagnetic field tensor is defined in terms of 4-vector potential as usual.
Invariance of action under infinitesimal coordinate translation leads to the 
covariant conservation law for the canonical energy (-momentum) tensor $T^{\mu\nu}$.
This canonical energy tensor is not symmetric and also not gauge invariant. Similar
to ${\bf r}\times{\bf p}$ the AM density tensor $M^{\mu\nu\lambda}$ can be
obtained from the canonical energy tensor, but it is found to be not conserved.  One can 
construct a symmetric, traceless, and gauge invariant energy tensor $E^{\mu\nu}$
such that
\begin{equation}
E^{\mu\nu} = T^{\mu\nu} + t^{\mu\nu}
\end{equation}
Here $t^{\mu\nu}$ is a divergenceless spin energy tensor. The AM density
tensor obtained from $E^{\mu\nu}$, let us denote it by $A^{\mu\nu\lambda}$ can be shown
to be conserved. Interestingly the AM density tensor $J^{\mu\nu\lambda}$
derived directly as a Noether current from the infinitesimal Lorentz rotation 
invariance of the action differs from $A^{\mu\nu\lambda}$, and the difference is a
pure divergence term so that the volume integrated AM in both cases is
identical. The lack of gauge 
invariance in the separation of spin and orbital parts is a well known problem; see van Enk and Nienhuis \cite{5} for recent discussion.This problem in classical 
field theory survives in quantum electrodynamics, and there arise intricate problems
in quantization of massless photon field. We wish to invite attention to another
crucial problem which does not find mention in the literature. Spin energy tensor
$t^{\mu\nu}$ does not contribute to the energy of the electromagnetic fields as a
consequence there is no rotational energy associated with radiation irrespective 
of its state of polarization. What does this signify?

Let us consider the expressions for energy and momentum densities given by
\begin{equation}
u = ({\bf E}^2 + {\bf B}^2)/8\pi
\end{equation}
\begin{equation}
{\bf g} =({\bf E}\times{\bf B})/4\pi c
\end{equation}

The time-averaged energy density and momentum density for monochromatic plane wave
satisfy the relation
\begin{equation}
u = |{\bf g}| c
\end{equation}
It is important to realize that this relation is independent of the state of polarization
of the plane wave: whole of the energy seems to be determined by the linear momentum,
and intrinsic spin has no energy. In a simple photon picture one relates energy E with 
momentum ${\bf p}$ by the relation
\begin{equation}
E = |{\bf p}| c
\end{equation}
This relation is analogous to the plane wave case i.e. Eq.(5). Using Planck's quantum
hypothesis for energy one can interpret photon momentum as $h\nu/c$ or
\begin{equation}
{\bf p} = \hbar {\bf k}
\end{equation}
Photon carries spin of $\hbar$ but energy of photon is determined solely by linear momentum.
In quantum optics polarization is taken into account using a polarization index in
the definition of the field operators, and once again there is no contribution to
the energy due to spin.

Beth experiment demonstrates the existence of SAM in the circularly
polarized light, therefore this problem cannot be set aside as a pseudo-problem.
For rigid body rotation about an axis of rotation in an inertial frame the
rotational kinetic energy essentially corresponds to the sum of the translational
kinetic energy of the constituent particles of the rigid body. The AM
in this case is defined to be extrinsic. In general there may exist pure rotation, 
Chasles' theorem says that a general displacement of a rigid body consists of pure
rotation plus translation \cite{16}. We argue that light and photon also possess both pure
rotation and translation, but there is a crucial difference, and it is more appropriate
to draw analogy with fluid motion for light \cite{4,17}. The problem of rotating
light could be approached following the insights gained in the analysis of rotating
superfluid. Feynman \cite{18} considers the rotation
of superfluid at absolute zero temperature, and concludes that vortex motion (in
fact, many vortices) is the lowest energy state. Note that vortex has important 
topological property of quantized circulation strength around the hole. The notion of
pure rotational energy that we suggest is the energy of the core, and the breakdown of 
continuum in the core region (though approximate for superfluid) endows topological property to both AM and its energy.

Preceding discussion leads us to picture photon as some kind of vortex with spin 
being a topological invariant. Polarized light could be thought of as comprising
of a large number of vortices, i.e. photons. Thus SAM is pure rotation and 
intrinsic and polarized light is a minimum energy state. For the OAM bearing beams
we distinguish two cases. In one case there exists a vortex-like rotation (e.g.
the LG beams) which has intrinsicness due to the vortices and their topological
property, but OAM depends on the reference axis of rotation hence it is extrinsic.
These could be treated as meta-stable energy states, and together with the AM spectrum considerations given in \cite{10}  their characterization as
quasi-intrinsic seems justified. In the second case the rotation is like a rigid 
body or rotation of the whole volume of the superfluid; OAM in this case is extrinsic.
Elliptical Gaussian beams would belong to this case, and these should be unstable
as compared to the LG beams. We suggest that studies on the far field behavior of
these beams may indicate their stability.

It becomes clear that the intrinsic nature of SAM implies that there must exist spin 
energy part in the total energy of the fields. A conservative approach is to accept the 
validity of the expressions (3) and (4), and ascribe part of
$u$ to pure rotational energy. The expressions for energy and momentum
densities calculated from the canonical energy tensor differ from the expressions
(3) and (4), and the difference is a pure divergence term for each one \cite{5}. On 
the other hand, symmetrical energy tensor gives the standard expressions. How to 
interpret spin energy in them? There are simple examples which show that
static magnetic field possesses AM. Usually for radiation only electric 
field part is calculated in $u$ and multiplied by 2 to get total energy. A tentative
suggestion is to identify magnetic energy part as rotational energy. The similarity
of action integral of rotating fluid and electromagnetic field has also been pointed
out in \cite{4}, and this also indicates that rotational energy resides in the 
magnetic field. 

Instead of the radiation fields it is more fruitful to reconsider the photon point of 
view. Historically energy $h\nu$ and momentum ${h\nu}/c$ were associated with light
quantum when the spin property of photon was unknown and continued to be
considered unimportant for many years. Assuming that spin is intrinsic pure rotation we
propose that the energy $h\nu$ is equally divided into pure rotational and translational parts
\begin{equation}
h\nu = (\hbar \omega)/2 + (h\nu)/2
\end{equation}

This splitting finds an interesting parallel with a classical particle having linear 
velocity $v$, momentum $p$, angular velocity $\omega$, and AM $L$.
Total kinetic energy for this particle is given by
\begin{equation}
K.E. = (L\omega)/2 + (p v)/2
\end{equation}
Let $v=c$, $p=h\nu/c$, and $L=\hbar$, then Eq.(9) goes over to Eq.(8). The parallel
drawn here should not be stretched too far; it is made in the spirit in which many
such analogies are made in the literature. For example, inertia is associated to
photon equating $mc^2$ with $h\nu$ to understand the phenomenon of the bending of 
light in general relativity. In fact the implication of the spin energy is on the
phase indeterminacy in the time dependent part of the phase (i.e. $\omega t$)
similar to the dislocation for the spatial phase. Cyclic internal time is a
topological obstruction in one dimensional line of time and would correspond to
topological torsion.

3. Spin as intrinsic or internal rotation implies that SAM to OAM conversion is
unphysical. How do we explain the experiments reported in
\cite{11}? Half-wave plate (HWP) of a uniaxial birefringent crystal converts 
polarization state of light; for example, a left circularly polarized plane wave $|L>$ at normal incidence emerges as right circularly polarized plane wave $|R>$ at the output port. A novel addition to HWP is made in this experiment: inhomogeneity is introduced orienting the fast (or slow) optical axis making an angle of $\alpha$ with the x-axis in the xy-plane for a planar slab
given by
\begin{equation}
\alpha(r,\phi) = q \phi + \alpha_0
\end{equation}
Here $r$ and $\phi$ are radial and azimuthal coordinates respectively, and z-axis is the direction of propagation of the assumed incident wave. Authors call this structure a q-plate. Jones calculus
applied at each point of the q-plate in a straightforward manner shows that the
output beam for an incident $|L>$ state is not only converted to $|R>$ state but it also 
acquires an azimuthal phase factor of $exp(2i q\phi)$. Similar to the LG beams
this phase is interpreted as OAM of $2q\hbar$ per photon in the output wave. Experiment
is carried out using nematic liquid crystal planar cell for $q=1$ and $TEM_{00}$ mode
Gaussian laser beam. Measurements on the interference pattern formed by the
superposition of the output beam with the reference beam display the wave front 
singularities and helical modes in the output beam.

The power of Jones calculus lies in the fact that properly applied it gives correct
formal results. However it leaves open the physical interpretation \cite{12}. The
optical process in q-plate and the experimental results obtained with $q=1$ plate
indicate that the physical mechanism responsible for the apparent spin to orbital
anglar momentum conversion is geometric phase in momentum or wave vector space.

Proposition: Geometric phase induced circular birefringence is the origin of 
topological phase singularity and OAM in q-plates.

We first briefly describe the spin redirection phase (SRP).
The unit wave vector ${\bf\kappa}$ and polarization vector $\epsilon ({\bf\kappa})$
describe the intrinsic properties of the light wave. The complex polarization vector
can be written in terms of two unit vectors ${\bf u}$, ${\bf v}$, and arbitrary phase 
between them. One can choose an orthonormal basis for the triad 
$({\bf\kappa},{\bf u},{\bf v})$. A plane wave propagating along a slowly varying path
in the real space, for example, in the coiled optical fibre, can be mapped on to
the surface of unit sphere in the wave vector space, and under a parallel transport along
a curve in this space preserving the spin helicity ${\bf\epsilon}.{\bf\kappa}$ the
polarization vector is found to be rotated after the completion of a closed cycle on the
sphere. The magnitude of the rotation is given by the solid angle enclosed by the
closed cycle, and the sign is determined by the initial polarization state. Since $|L>$
and $|R>$ states acquire equal but opposite geometric phases, Berry has aptly termed
this effect as geometric circular birefringence \cite{14}.

In the Marrucci et al experiment the inhomogeneity is in the polarization in the
transverse plane, therefore it may not be immediately obvious to envisage the role of 
wave vector parameter space. To understand this we recall the geometric approach of Nye 
and Hajnal in \cite{7}. Electric field rotation is represented by a general polarization 
ellipse but the direction of propagation is a subtle issue in the presence 
of singularities of the vector wavefield. Decomposing the elliptical polarization 
into the superposition of circular polarizations with amplitudes $(\rho_1,\rho_2)$ and wave vectors $({\bf k}_1,{\bf k}_2)$ a weighted mean $({\bf k}_w)$ of the two is defined for the 
propagation direction. In the case of circular polarization the difference 
between the two wave vectors ${\bf K}$ determines the index of singularity 
${\pm 1/2}$, and the "subordinate" circular state gives the vortex-like phase 
dislocation for which $\rho_2$ is zero but the phase is indeterminate. We picture equivalent description in which helicity preserving transformation occurs in the weighted wave vector space. Since the polarization variation is confined in the transverse plane for the q-plates the constraint of the spin helicity
${\bf\epsilon}.{\bf\kappa}$ preserving process in the q-plate with the azimuthal
dependence of $\alpha$ given by Eq.(10) would lead to spiral path for the wave vector.
In q=1 plate the circular plus linear propagation along z-axis will result into a
helical path and the width of the plate (HWP) ensures that the input and output
ends of the helix are parallel. On the unit sphere in the wave vector space this
will correspond to a great circle, and the solid angle would be $2\pi$. Since the
SRP equals the solid angle this result is in agreement with Eq.(3)
of \cite{11}. The important observation emphasized by Marrucci et al that the incident
polarization controls the sign of the orbital helicity or topological charge is easily
explained in view of the property of the geometric birefingence in which handedness
decides the sign of the phase. Thus both magnitude and sign of the azimuthal phase 
have been explained in accordance with the proposition.

Let us analyze the claim of SAM to OAM conversion made by the authors. The q=1 plate
is a special optical element in which no transfer of AM to the crystal 
takes place, and total AM is conserved within the light beam. We argue
that spin is intrinsic, and the OAM is a manifestation of the geometric phase with exact
equivalence between them in this case. In our interpretation the AM holonomy is related with the geometric phase \cite{12,13}. Thus Marrucci
et al experiment seems to offer first direct evidence in support of our conjecture.

Towards the end of their paper authors have alluded to the role of the Pancharatnam-
Berry phase following \cite{19}. The experiments performed using space-variant gratings
in \cite{19} also show the appearance of OAM; our arguments extended to these experiments
imply that the SRP rather than the Pancharatnam phase should occur here.
Perhaps a possible reason for invoking the Pancharatnam phase in \cite{19} could be the
idea of the Pancharatnam connection used by Nye \cite{7} in the later exposition of
singularities. Note that Pancharatnam connection defines the phase difference $(\delta)$
between two arbitrary nonorthogonal polarization states in terms of the interference
formula. Nye uses this definition to arrive at a wave vector ${\bf k}_\delta$ which
is exactly same as the weighted wave vector of Nye and Hajnal. However geometric
phase on the Poincare sphere with fixed direction of propagation is not involved
in the wavefront dislocation.

We suggest further experiments are necessary to clarify this issue, and also to test the
AM holonomy conjecture unambiguously. The circularly polarized
light when passed through a combination of two HWPs one of which is made to rotate slowly
though restores its polarization but gets frequency shifted, and the shift could be interpreted \cite{20} in terms of the evolving geometric phase on the Poincare sphere.
Analogous frequency-shift experiments for OAM bearing beams are reviewed in Allen et al
\cite{6}. Additive property of SRP and Pancharatnam phase is also
established in the literature. A logical and simple novel modification is to carry out
rotating q-plate experiments; such experiments may throw light on the role of geometric
phases in vector vortices, and also delineate the spin-orbit coupling, if any. More
rewarding experiments might be at a single photon level. For rigorous analysis we suggest
Jones calculus for the Gaussian beams \cite{21} may be useful.

4. To conclude the Letter two contributions on the problem of AM of light
in the current perspective have been made. The first concerns with the debate on the
intrinsicness or otherwise of OAM. We argue that the rotational energy is crucial
for defining the intrinsic or extrinsic property of light. Hitherto neglected obscure
nature of spin energy is discussed, and a plausible suggestion is made that magnetic 
energy part represents the spin energy. Photon is proposed to possess
linear momentum $h\nu/c$ and corresponding translational energy $h\nu/2$, and spin ${\pm\hbar}$ and corresponding pure rotational energy $\hbar\omega/2$. Spin of photon is intrinsic as being a topological invariant and the spin energy is due to topological torsion. The OAM for 
elliptical Gaussian beams is pure extrinsic while LG beams have quasi-intrinsic OAM due to vortices.

The issue  of spin to orbital conversion \cite{11} is resolved in terms of SRP:
geometric phase induced circular birefringence is facilitated by the crystal
optics of q=1 plate resulting into the observed azimuthal phase structure. The exact
equivalence of the OAM with the geometric phase in this special case seems to offer
first direct support to our conjecture on AM holonomy \cite{12,13}. Note that
recently another experiment \cite{22} has also shown consistency with our conjecture. Geometric phase and AM of light hold promise for further deeper understanding on the nature of light
and crystal optics.

I am grateful to Prof L. Allen for explaining me the meaning of quasi-intrinsic 
nature and reference \cite{9}. The Library facility at Banaras Hindu University
is acknowledged.

\end{document}